\documentclass{aa}  

\usepackage{graphicx,natbib,subcaption}
\usepackage{txfonts}
\usepackage[]{hyperref}
\begin{document}

\title{
Nebular phase observations of the type-Ib supernova iPTF13bvn favour a binary progenitor
\thanks{Based on observations obtained at the Southern Astrophysical Research (SOAR) telescope, which is a joint project of the Minist\'{e}rio da Ci\^{e}ncia, Tecnologia, e Inova\c{c}\~{a}o (MCTI) da Rep\'{u}blica Federativa do Brasil, the U.S. National Optical Astronomy Observatory (NOAO), the University of North Carolina at Chapel Hill (UNC), and Michigan State University (MSU); Chilean Telescope Time Allocation Committee proposal CN2014A-91.}
}


   \author{
H. Kuncarayakti\inst{1,2}\thanks{\email{hanin@das.uchile.cl}}\and
K. Maeda\inst{3,4}\and
M.~C. Bersten\inst{5,4}\and
G. Folatelli\inst{5,4}\and
N. Morrell\inst{6}\and
E.~Y. Hsiao\inst{6}\and
S. Gonz\'alez-Gait\'an\inst{1,2}\and
J.~P. Anderson\inst{7}\and
M. Hamuy\inst{2,1}\and
T. de Jaeger\inst{1,2}\and
C.~P. Guti\'errez\inst{1,2,7}\and
K.~S. Kawabata\inst{8, 9}
          }

   \institute{Millennium Institute of Astrophysics, Casilla 36-D, Santiago, Chile
        \and
             Departamento de Astronom\'ia, Universidad de Chile, Casilla 36-D, Santiago, Chile
             \and
             Department of Astronomy, Kyoto University, Kitashirakawa-Oiwake-cho, Sakyo-ku, Kyoto 606-8502, Japan 
             \and
             Kavli Institute for the Physics and Mathematics of the Universe (WPI), The University of Tokyo, Kashiwa, Chiba 277-8583, Japan
             \and
             Instituto de Astrofísica de La Plata (IALP), CCT-CONICET-UNLP, Paseo del Bosque S/N, B1900FWA La Plata, Argentina
             \and
             Las Campanas Observatory, Carnegie Observatories, Casilla 601, La Serena, Chile
             \and
             European Southern Observatory, Alonso de C\'ordova 3107, Vitacura, Santiago, Chile
             \and
             Hiroshima Astrophysical Science Center, Hiroshima University, Higashi-Hiroshima, Hiroshima 739-8526, Japan
			\and
			Core of Research for the Energetic Universe, Hiroshima University, Higashi-Hiroshima, Hiroshima 739-8526, Japan
             }

   \date{Received XXXXX; accepted XXXXX}

 
  \abstract
  {}
   {We present and analyse late-time observations of the type-Ib supernova with possible pre-supernova progenitor detection, iPTF13bvn, taken at $\sim$300 days after the explosion, and discuss these in the context of constraints on the supernova's progenitor. Previous studies have proposed two possible natures for the progenitor of the supernova, i.e. a massive Wolf-Rayet star or a lower-mass star in close binary system.}
  {Our observations {show} that the supernova has entered the nebular phase, with the spectrum dominated by Mg~I]$\lambda\lambda$4571, [O~I]$\lambda\lambda$6300, 6364, and [Ca~II]$\lambda\lambda${{7291}}, 7324 emission lines. We measured the emission line fluxes to estimate the core oxygen mass and compare the [O~I]/[Ca~II] line ratio with other supernovae. }
   {
{   The core oxygen mass of the supernova progenitor was estimated to be $\lesssim$0.7 M$_\odot$, which implies initial progenitor mass not exceeding $\sim$15 -- 17 M$_\odot$. }
   Since the derived mass is too small for a single star to become a Wolf-Rayet star, this result lends more support to the binary nature of the progenitor star of iPTF13bvn.
   The comparison of [O~I]/[Ca~II] line ratio with other supernovae also shows that iPTF13bvn appears to be in close association with the lower-mass progenitors of stripped-envelope and type-II supernovae.
   }
   {}



   \maketitle
%

\section{Introduction}
Supernovae (SNe) type Ib and Ic, which do not show hydrogen lines in their spectra, are thought to be produced by the core-collapse of massive stars which have lost their outer hydrogen envelope. Hydrogen-deficient Wolf-Rayet (WR) stars have been proposed as the possible progenitors for these SNe \citep{maeder82,ensman88}. However, thus far there have been no strong confirmation of WR star as SN Ib/c progenitor. As summarized by \citet{eldridge13}, examinations of \textit{Hubble Space Telescope (HST)} pre-SN archival images of SNe Ib/c in nearby galaxies has not succeeded in detecting the progenitor stars. 
{\citet{yoon12} argued that type-Ib/c SN progenitors are faint in optical bands during the pre-SN stage thus direct detections have not been providing strong constraints on the progenitor luminosity and mass.}

\citet{cao13} reported the discovery and early evolution of \object{iPTF13bvn}, a type-Ib SN discovered by the intermediate Palomar Transient Factory (iPTF) survey \citep{rau09,law09} in the host galaxy \object{NGC 5806}. It was found in the \textit{HST} archives that the SN position coincides with a blue source, within 2$\sigma$ (80 mas, $\approx$ 8.7 pc) positional uncertainty of the SN. The photometry of the object was shown to be consistent with a WR star, which was suggested as the progenitor of iPTF13bvn. 
\citet{groh13} proposed that their model of WR star with initial mass of 31 -- 35 M$_\odot$ could reproduce the observed photometry of the object, the candidate progenitor star of iPTF13bvn. 
If the disappearance of the candidate progenitor star is confirmed by future observations, this would be the first detection of a progenitor of a type-Ib SN.

An alternative scenario was proposed by \citet{bersten14}, in which the progenitor star was modeled to be in a close binary system. Their hydrodynamic model shows that the primary star in this system evolved and lost its hydrogen layer due to binary interaction, and exploded as a $\approx$3.5 M$_\odot$ pre-SN star, corresponding to the progenitor initial mass of {20} M$_\odot$. This model successfully reproduced the observed light curve, absence of hydrogen, and photometry of the pre-SN candidate progenitor star. The notion that the progenitor star of iPTF13bvn could not have been a massive WR star, but instead a lower-mass star in binary system, has also been suggested by \citet{fremling14} and \citet{srivastav14}. 
\citet{eldridge15} argued that the photometry of \citet{cao13} underestimated the brightness of the candidate progenitor object by up to 0.7 mag, which brought them into the conclusion that the observed spectral energy distribution does not match a typical single WR star, but rather more consistent with their binary models.

In this paper we present optical photometry and spectroscopy of iPTF13bvn obtained during its nebular phase, at 306 days after the explosion date, from which we derive an estimate of the initial mass of the SN progenitor star.

%
%


\section{Observations and data reduction}

The observations of iPTF13bvn {were} carried out using the 4.1 m Southern Astrophysical Research (SOAR) telescope stationed at Cerro Pach\'on, Chile at an altitude of 2700 m above sea level, starting in the night of 18 April 2014 (Chilean local time) in visitor mode. This observation date corresponds to $\sim$306 days after the assumed explosion time of JD = 2456459.24$\pm$0.9 \citep{bersten14}, or +290 days past the time of maximum light.

\begin{figure*}
\centering
\includegraphics[width=\linewidth]{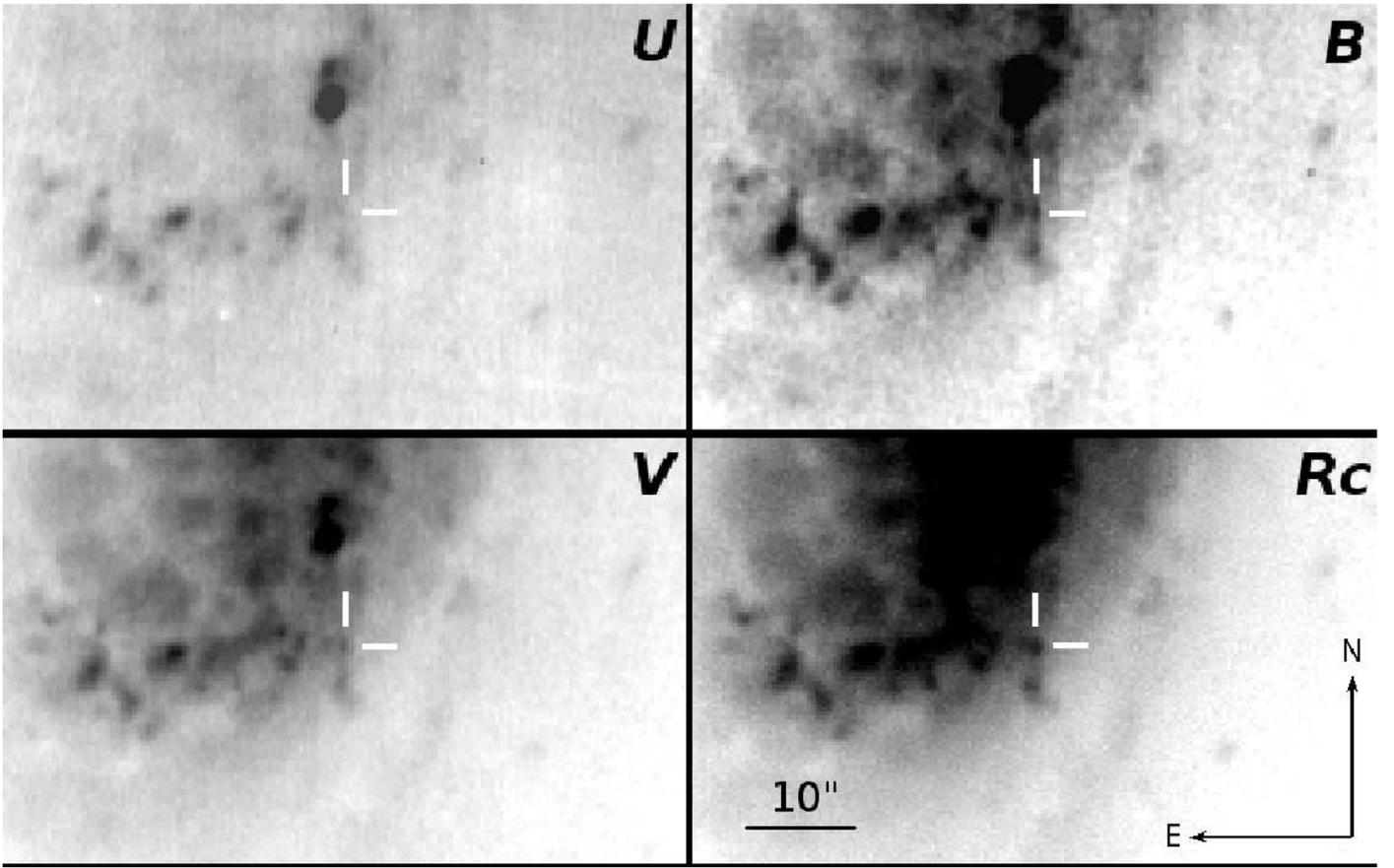}
\caption{Sections of \textit{UBVRc} images of iPTF13bvn, with the SN indicated.\label{image}}
\end{figure*}

We used the Goodman High Throughput Spectrograph \citep{clemens04} for both photometry and spectroscopy. The Goodman HTS employs a Fairchild 4k$\times$4k CCD with a physical scale of 0.15"/pixel. The unvignetted field of view size of the instrument is around 7.2' in diameter.
We used the 2$\times$2 binning mode for both photometry and spectroscopy. During the night, the sky was clear but not photometric, with variable seeing. Table~\ref{tbllog} shows the observation log for the night. The observation of iPTF13bvn was started with spectroscopy, followed by photometry. Spectrophotometric standard star and photometric standard fields \citep{landolt92} were also observed during the night.

%
   \begin{table}
      \caption[]{Observing log.}
         \label{tbllog}
     $$ 
         \begin{array}{lrcc}
            \hline
            \noalign{\smallskip}
            \textrm{Object}  &  \textrm{Exp. time (s)}  & \textrm{Airmass} & \textrm{Seeing (")}\\
            \noalign{\smallskip}
            \hline
            \noalign{\smallskip}
            \textrm{iPTF13bvn spec.} & \textrm{5100 } & 1.4-1.2 & 1.2-0.9 \\ 
            \textrm{HR5501 spec.} & \textrm{1 } & 1.2 & 0.9 \\ 
            \textrm{iPTF13bvn } BVR_c & \textrm{300 } & 1.2-1.3 & 0.9-1.2 \\ 
            \textrm{iPTF13bvn } U & \textrm{1800 } & 1.3 & 1.2 \\ 
            \textrm{PG1528 } BVR_c & 1-3 \textrm{ } & 1.4 & 1.3 \\ 
            \textrm{PG1528 } U & 30 \textrm{ } & 1.4 & 1.3 \\ 
            \textrm{PG1525 } BVR_c & 2-5 \textrm{ } & 1.3 & 1.2 \\ 
            \textrm{PG1525 } U & 30 \textrm{ } & 1.3 & 1.2 \\             
            \noalign{\smallskip}
            \hline
         \end{array}
     $$ 
  \textsc{Note:} HR5501 is a spectrophotometric standard star; PG1525 and PG1528 are Landolt photometric standard fields.
   \end{table}
%
%

We used the 300 lines/mm grating with GG385 blocking filter and 1.03" slit mask, positioned at {the} parallactic angle, for our spectroscopic observation of iPTF13bvn. With this instrument configuration, the dispersion for 2$\times$2 pixel binning is 2.6 \AA/pixel with wavelength coverage 3600 -- 8800 \AA. We took a 1500 s exposure and 2$\times$1800 s exposures of iPTF13bvn, resulting in a total of 5100 s exposure time for the object. We used the 3" slit for the spectrophotometric standard star.
An internal HgAr arc lamp was used for wavelength calibration, and internal quartz lamp was used for spectral flat calibration. 
{From the arc spectra we measured the effective full width at half maximum (FWHM) resolution of the spectra to be $\approx$7.6 \AA~ around the wavelength 7000 \AA.}
The raw data was reduced using \textsc{Iraf}\footnote{\textsc{IRAF} is distributed by the National Optical Astronomy Observatory, which is operated by the Association of Universities for Research in Astronomy (AURA) under cooperative agreement with the National Science Foundation.} in a standard slit-spectroscopy reduction manner, with the \textsc{twodspec} and \textsc{onedspec} packages. Wavelength and flux calibrations were also performed. 

Photometry was carried out in \textit{UBVRc} bands. Figure~\ref{image} shows the position of the SN within its host galaxy. The SN is clearly detected in the \textit{Rc} band, as well as in \textit{B} and \textit{V} although not as clearly. In the \textit{U}-band, the SN is not detected within 2$\sigma$ of the background noise. The integration time was 300 s for the \textit{BVRc} filters, and 1800 s for the \textit{U} filter.
Point-spread function (PSF) photometry was performed to the \textit{BVRc} images and aperture photometry was done to the \textit{U}-band image using aperture value equal to the seeing size, after standard image reduction procedures with \textsc{Iraf}. PSF photometry was performed using the \textsc{daophot} package \citep{stetson87}, and aperture photometry using \textsc{apphot}.

\section{Results and discussions}

\subsection{Spectroscopy}

\begin{figure*}[]
\centering
\includegraphics[width=\linewidth]{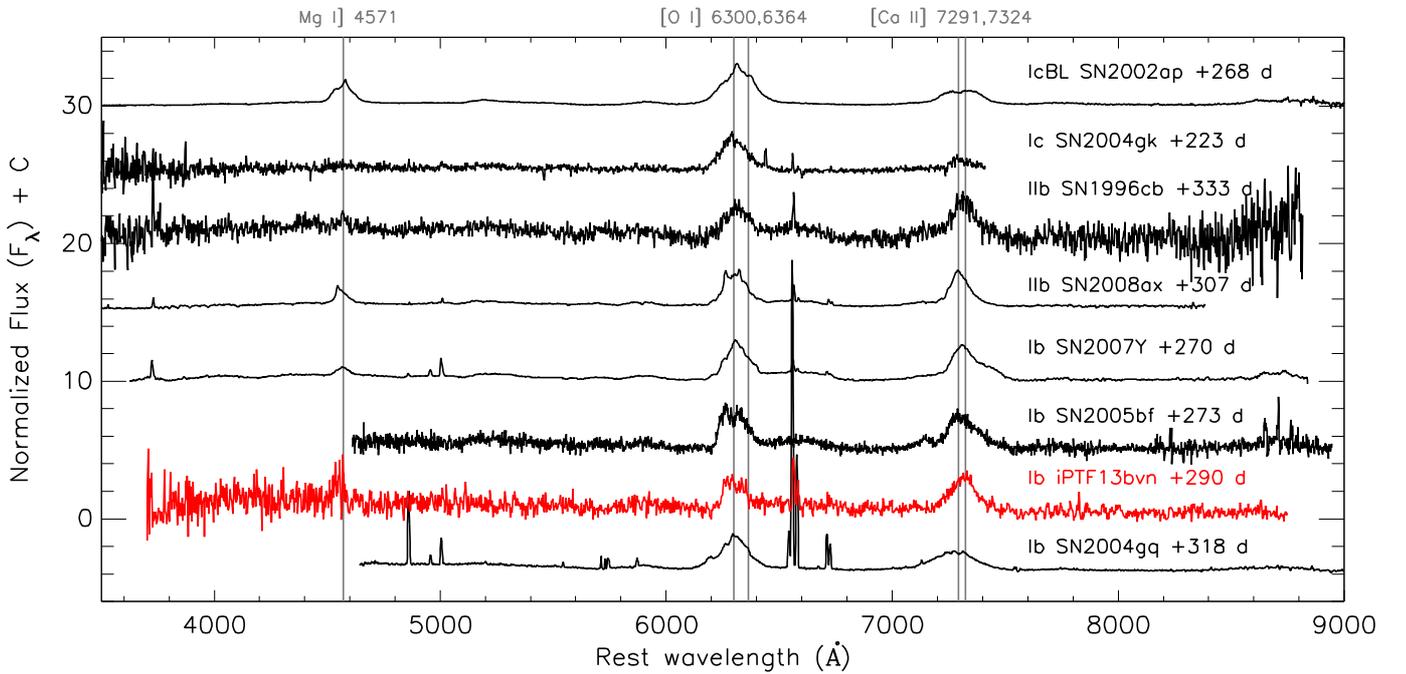}
\caption{{Spectrum of iPTF13bvn at 306 days past explosion (+290 days after maximum light), compared to several stripped-envelope core-collapse SNe at similar ages from \citet[][]{modjaz14}, \citet[][SN 2005bf]{maeda07}, and \citet[][SN 2004gq]{maeda08}. }
The phase of the spectra in days are with respect to the time of maximum light.
The prominent emission lines originated from the SNe are indicated with vertical grey lines. The spectra are normalized to the peak of the [O I] line of iPTF13bvn and shifted in flux for presentation purposes. The narrow emission lines originate from the underlying H II regions. 
\label{specs}}
\end{figure*}

Our observation was conducted at 306 days post SN explosion. At this age, the SN is expected to have already entered the nebular phase, which is confirmed by the observed spectrum (Figure~\ref{specs}). The continuum of the spectrum is very weak, with strong emission lines of [O I]$\lambda\lambda$6300, 6364 and [Ca II]$\lambda\lambda${{7291}}, 7324 dominating. In addition to the [O I] and [Ca II] lines, the spectrum also exhibits a noticeable Mg I]$\lambda4571$ semi-forbidden intercombination line. These spectral features are commonly shown by stripped-envelope core-collapse SNe at nebular stage (see Figure~\ref{specs}). 
{Nebular spectra of seven other SNe \citep{modjaz14,maeda07,maeda08} are shown alongside iPTF13bvn for comparison in Figure~\ref{specs}.}

The nebular spectrum of iPTF13bvn looks remarkably similar to the type-IIb SN 2008ax at +307 days post-maximum, despite the {noticeably weaker} signal-to-noise ratio. SNe IIb are thought to be closely related to SNe Ib -- both subclasses show prominent helium lines, but the hydrogen lines present in type-IIb SNe are observed only during the early phase. Thus, it is possible that many SNe IIb might have been misidentified as a type-Ib. Indeed it has been suggested that the IIb or Ib classification may be time-dependent \citep{mil13}. In the case of iPTF13bvn, hydrogen was not detected in spectra as early as +2.3 days after the explosion, hence the Ib classification \citep{cao13}{; however, see the discussions about possible H$\alpha$ emission in subsection \ref{sec:ha}.}

\subsubsection{The [O I] line profile}
\begin{figure}[]
\centering
\includegraphics[width=\linewidth]{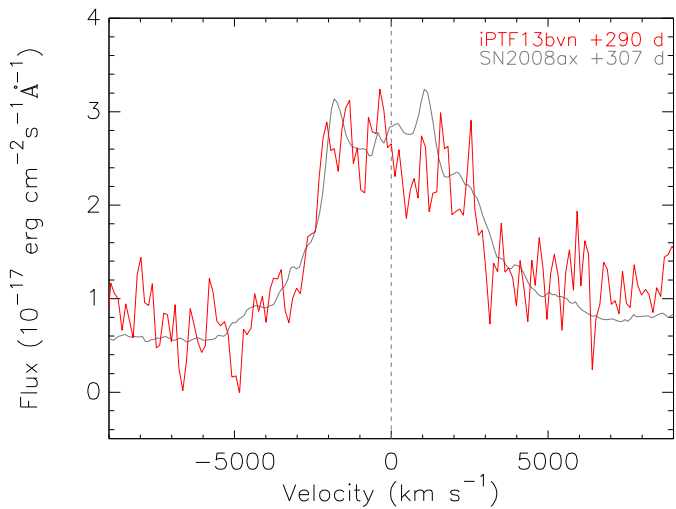}
\caption{Line profile of [O III]$\lambda\lambda$6300, 6364 of iPTF13bvn compared to SN 2008ax. The dashed vertical line corresponds to zero velocity with respect to 6300 \AA. The flux of the spectrum of SN 2008ax has been normalized and shifted. \label{line6300}}
\end{figure}

The [O I]$\lambda\lambda$6300, 6364 emission line of iPTF13bvn evidently does not show a single-peak, Gaussian-like profile. It exhibits a rather boxy, flat-topped profile, that to some extent is suggestive of a double-peaked emission. In Figure~\ref{line6300} we compare the line profile of the [O I]$\lambda\lambda$6300, 6364 emission in iPTF13bvn to that of SN 2008ax at the same phase. The spectrum of SN 2008ax is the same as the one presented in Figure~\ref{specs}. Both spectra show [O~I] FWHM of $\sim$5000 km~s$^{-1}$. While the [Ca~II] line of iPTF13bvn also show a similar velocity FWHM of $\sim$5000 km~s$^{-1}$, in the case of SN 2008ax the FWHM is lower, $\sim$4000 km~s$^{-1}$. These are within the likely core velocity values of $\sim$ 1000 -- 5000 km~s$^{-1}$ as suggested by \citet{fransson89}.

\begin{figure}[]
\centering
\includegraphics[width=\linewidth]{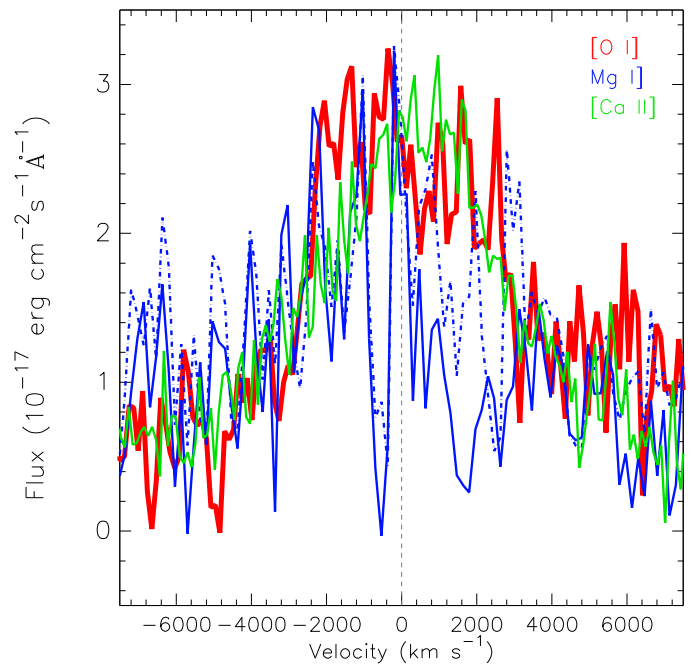}
\caption{
{Line profiles of [O~I], [Ca~II], and Mg~I], in velocity space. Zero velocities are with respect to 6300 \AA, 7308 \AA, and 4571 \AA, respectively. The fluxes of [Ca~II] and Mg~I] lines are normalized to the [O~I] line. The synthetic Mg~I] line profile is represented with dash-dot line (see text for details).
} 
\label{OMgCa}}
\end{figure}

{
In Figure~\ref{OMgCa} we present the profiles of the [O~I], [Ca~II], and Mg~I] emission lines. Here we plot the observed line profile of Mg~I] with full line, and additionally a synthetic line profile with dash-dot line. This synthetic profile was generated to simulate a doublet feature for Mg~I], to allow direct comparison with the [O~I] doublet. We follow the method described in \citet{taubenberger09}, in which the Mg~I] line is scaled down to 1/3 of its initial intensity then shifted by +46 \AA{} and added to the original spectrum to generate the synthetic doublet. \citet{taubenberger09} found that for most objects in their sample ($\sim65\%$) the [O~I] and Mg~I] lines show remarkably similar profiles. It is interesting that even with a synthetic doublet, the red wing of the Mg~I] line in iPTF13bvn seems to be depressed in comparison with the profiles of [O~I] and [Ca~II] although there is the caveat of low signal-to-noise ratio. The strength of this line is also considerably stronger compared to some other SNe as apparent in Figure~\ref{specs}. 
The Mg~I]/[O~I] flux ratio in iPTF13bvn is around 0.85, whereas for SNe 2002ap, 2008ax, and 2007Y the value is around 0.2 -- 0.3. \citet{foley03} noticed the exceptional strength of this Mg~I] line in SN 2002ap at late times and that it is growing with time compared to the strength of [O~I] and [Ca~II] lines. As time passes we are seeing deeper into the progenitor core and reach the Mg-O layer, thus it is expected that this line grows stronger at later times.
For the [Ca~II] line in iPTF13bvn, the line profile appears to be less asymmetric compared to the other two lines. These might indicate different geometry and distribution of the emitting materials for [O~I], [Ca~II], and [Mg~I].
}

The profile shape of the [O~I] line in iPTF13bvn suggests that there is some degree of asphericity in the ejecta. While it is not very clearly double-peaked, the line profile is considerably different to the more Gaussian-like [Ca~II] line. \citet{maeda08} showed that double-peaked [O~I]$\lambda\lambda$6300, 6364 emission commonly observed in late-time spectra of core-collapse SNe signifies asphericity in the SN explosion. \citet{tanaka09} also reported that the [O~I] line in the nebular spectrum of the type-Ib SN 2008D can be explained by a bipolar explosion with a torus-like distribution of oxygen.
\citet{taubenberger11}, who also observed the similar feature in SN 2008ax, argued that a spherical oxygen distribution with a significant clump or torus, or even an aspherical distribution of $^{56}$Ni that excites the oxygen, may explain the observed feature. 
{On the other hand, the asymmetry in the line profile could be attributed to absorption in the interior, which may be caused by dust, or high opacity associated with clumping or high density material \citep[see e.g.][]{taubenberger09,mil10}.
In this case, the observed double-peaked profile may also not necessarily reflect a torus or elongated-shell geometry of the oxygen emitting region, but simply the doublet nature of the line as suggested by \citet{mil10}. Although it cannot be ruled out in all cases, the doublet explanation cannot account for the observed  ratios of the components, nor for the cases where the profile is singly peaked.
}

\subsubsection{Oxygen and progenitor mass}
\label{sec:oxmass}
Using the flux of the [O~I]$\lambda\lambda$6300, 6364 emission line, the mass of oxygen that produced the line can be estimated. \citet{jerkstrand12} have demonstrated that the strength of the [O~I]$\lambda\lambda$6300, 6364 line is sensitive to the variation of the initial mass of the SN progenitor star.

{\citet{uomoto86} provided an equation to calculate the minimum oxygen mass (in M$_\odot$ unit) required to produce the emission, as follows:}
\begin{equation}
M_{oxygen} = 10^8 f([\textrm{O I}]) D^2 \exp(2.28/T_4).
\label{equ}
\end{equation}

Therefore, using this equation one can calculate the oxygen mass from the measured flux of the [O~I]$\lambda\lambda$6300, 6364 line in erg s$^{-1}$ cm$^{-2}$, $f([\textrm{O~I}])$, assuming that the distance in Mpc $D$ and temperature of the oxygen-emitting region $T_4$, in $10^4$ K units, are known. 
{The distance to the host galaxy of iPTF13bvn NGC 5806 derived with Tully-Fisher method is known to be $\sim$26 Mpc. The Extragalactic Distance Database\footnote{\url{http://edd.ifa.hawaii.edu/}} \citep{tully09} gives the distance to NGC 5806 as $26.2\pm1.5$ Mpc, and this has been revised with a newer value of $26.79\pm0.2$ Mpc in the Cosmicflows-2 catalog \citep{tully13}. We adopt this newer value as the distance to iPTF13bvn. For reference, \citet{cao13} and \citet{srivastav14} adopt the distance of 22.5 Mpc, while \citet{bersten14} uses 25.5 Mpc. From Equation \ref{equ} it is clear that adopting smaller distance would decrease the estimate of the oxygen mass, therefore the progenitor mass.
}

{The measurement of the emission line flux of [O~I]$\lambda\lambda$6300, 6364 was done using the task \texttt{splot} in \textsc{Iraf} with three different methods. Prior to the flux measurements, we corrected the spectrum of iPTF13bvn for Milky Way and host galaxy extinctions, amounting $E(B-V)_{MW} = 0.0447$ mag and $E(B-V)_{host} = 0.17$ mag \citep{bersten14}, assuming a standard \citet{cardelli89} interstellar reddening law with $R_V = 3.1$, and removed the spectral continuum by fitting a first-order cubic spline function. Measurement of the line flux by direct integration gives $3.46\times10^{-15}$ erg s$^{-1}$ cm$^{-2}$, by fitting a single gaussian function $3.96\times10^{-15}$ erg s$^{-1}$ cm$^{-2}$, and by deblending the profile into two gaussians $3.70\times10^{-15}$ erg s$^{-1}$ cm$^{-2}$. We use the average of the three measurements, $3.71\pm0.25\times10^{-15}$ erg s$^{-1}$ cm$^{-2}$ as the flux of the oxygen line, to be used in equation~\ref{equ}.
A possibly significant source of uncertainty in this method of \citet{uomoto86} is the time evolution of the strength of the [O~I] line as well as the temperature $T_4$. }

One way to estimate $T_4$ is by measuring the ratio of [O~I]$\lambda5577$ line to [O~I]$\lambda\lambda$6300, 6364. However, in the case of iPTF13bvn we could not detect this weak line at 5577 \AA, which is actually an indication of low temperature \citep{elmhamdi11}. 
The near-infrared Ca~II line was also not detected in the iPTF13bvn spectrum, further indicating the low temperature.
\citet{sollerman98} used this method of [O~I]$\lambda5577$ line to estimate the temperature to be $<$ 5000 -- 4400 K, which in turn gives an oxygen mass of 0.11 -- 0.21 M$_\odot$ for SN Ib 1996N. \citet{sahu11}, not detecting the [O~I]$\lambda5577$ in the nebular spectra of SN Ib 2009jf, used $T_4 = 0.4$ to derive oxygen mass of 1.34 M$_\odot$. \citet{elmhamdi04}, using the upper limit for the [O~I]$\lambda5577$ line, derived oxygen mass of 0.7 -- 1.35 M$_\odot$ for SN Ib 1990I, for temperatures between 3200 -- 3500 K {at +237 days}. 

{Assuming the lower limit of the temperature of the oxygen-emitting region to be 3200 K as in SN 1990I, while adopting the line flux $3.71\pm0.25\times10^{-15}$ erg s$^{-1}$ cm$^{-2}$ and distance $26.79\pm0.2$ Mpc, we calculated the minimum oxygen mass in iPTF13bvn using Equation~(\ref{equ}) to be $0.33\pm0.03$~M$_\odot$.
}
This oxygen mass value is relatively low compared to SNe 1990I and 2009jf, but comparable to SN 1996N if the oxygen temperature of iPTF13bvn is assumed to be slightly higher. With higher temperature, Equation~(\ref{equ}) will yield an even lower value of oxygen mass.
Again here we note the sensitivity of Equation~(\ref{equ}) to the temperature assumption -- increasing the temperature from 3200 K to 5000 K will drop the derived oxygen mass by a factor of $\sim$13, while on the opposite direction using very cool temperature less than 2500 K will increase the derived oxygen mass by one order of magnitude. At any rate, a temperature less than $\sim$3000 K is not plausible as beyond this value the cooling will shift to be dominated by far-infrared fine-structure rather than optical lines, resulting in a thermal instability and sudden temperature drop \citep{fransson89}.
{Thus, it is not likely that the minimum oxygen mass is significantly higher than the derived value of 0.33 M$_\odot$}

{It is to be noted, however, that Equation~\ref{equ} provides the minimum oxygen mass responsible for the line emission, which may not be equal to the total oxygen mass. The presence of non-optically thin materials and clumping is not accounted for, thus the total oxygen mass may actually be higher than the derived value of 0.33 M$_\odot$. \citet{maeda06aj} derived a total oxygen mass of 2.3 M$_\odot$ for SN 2006aj, 1.3 M$_\odot$ of which is within $v\lesssim7300$ km s$^{-1}$ and consistent with the luminosity of the [O~I]$\lambda\lambda$6300, 6364 lines. With a conservative assumption that $\sim50\%$ of the total oxygen mass in iPTF13bvn is responsible for the emission line, then the total oxygen mass should not exceed $\sim$0.7 M$_\odot$. The effects of temperature assumption and non-emissive oxygen seem to dominate the uncertainty in the oxygen mass derivation, comparing with the uncertainties in distance or flux measurement. 
}

\begin{figure}[]
\centering
\includegraphics[width=\linewidth]{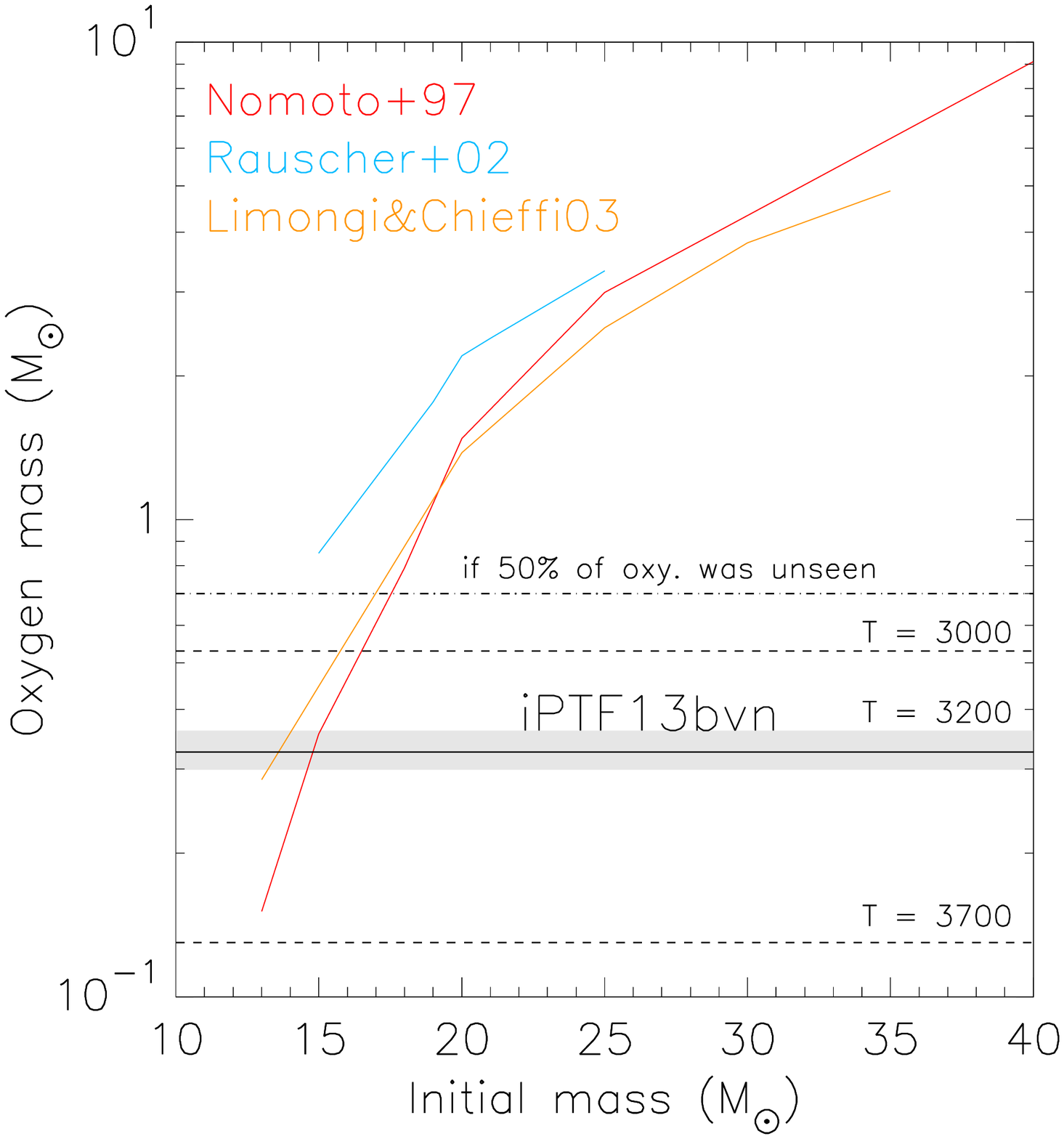}
\caption{
{Oxygen yield of the explosion of massive stars of various main sequence masses ({\citealt{nomoto97}: red, \citealt{rauscher02}: light blue, \citealt{limongi03}: orange}). Horizontal line indicates the estimated oxygen mass for iPTF13bvn assuming $T = 3200$~K, with uncertainty represented by shaded region. Dashed horizontal lines denote the oxygen mass values for different temperature assumptions. The estimated oxygen mass if 50$\%$ of oxygen was unseen is indicated with dash-dot line.
}\label{oxmass}}
\end{figure}

In Figure~\ref{oxmass} we compare the derived oxygen mass with nucleosynthesis oxygen yields of massive stars of various initial masses, from \citet{nomoto97}, \citet{rauscher02}, and \citet{limongi03}. 
{It is apparent that the derived oxygen mass of iPTF13bvn is more consistent with what would be produced from the explosion of a massive star with mass $\lesssim15$ -- 17 M$_\odot$, even if the total oxygen mass is assumed to be twice the derived value from the [O~I]$\lambda\lambda$6300, 6364 line. In an extreme case in which the whole SN ejecta \citep[$\lesssim2.3$ M$_\odot$;][]{bersten14,srivastav14} were assumed to be oxygen, which is not realistic, the initial progenitor mass of iPTF13bvn would still be smaller than $\sim$25 M$_\odot$. For reference, the derived oxygen masses of SNe 1990I and 2009jf are around 14 -- 36 $\%$ of the total ejecta mass \citep{elmhamdi04,sahu11}.}
This implies that a single Wolf-Rayet progenitor scenario for iPTF13bvn, whose initial mass must have been $\gtrsim25$ M$_\odot$ at solar metallicity \citep{crowther07}, is unlikely. 
Indeed, a lower-mass binary progenitor scenario for iPTF13bvn seems to be more plausible \citep{bersten14,fremling14,srivastav14,eldridge15}.
With lower metallicity, the WR mass limit would move to even higher mass due to the reduced strength of the metallicity-driven wind. In this context, the metallicity of the explosion site of iPTF13bvn was measured to be slightly less than solar (subsection~\ref{sec:metal}). 
{We note that nucleosynthesis yields are possibly subject to uncertainties introduced by neglecting the effects of turbulence in the interior \citep{smith14}, hence this may affect the estimate of the initial mass of the SN progenitor star. However, this effect is still not thoroughly studied and quantified.
}

\begin{figure}[]
\centering
\includegraphics[width=\linewidth]{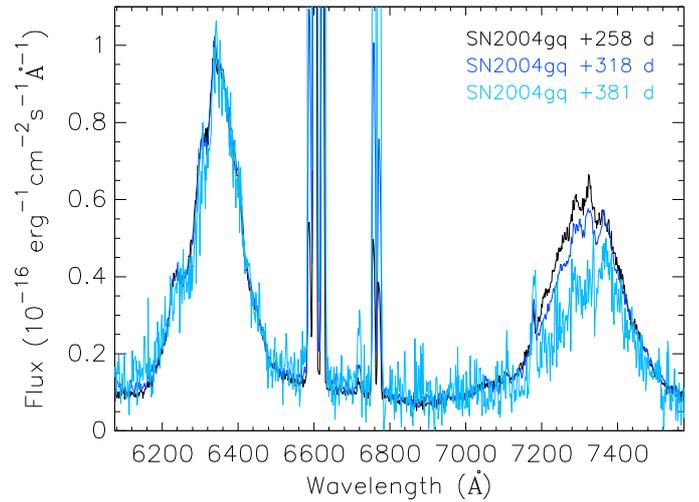}
\caption{
{Evolution of the nebular spectrum of SN 2004gq around the [O I] -- [Ca II] spectral region, between +258 and +381 days post maximum. Spectra has been scaled to match the [O I ] line at +258 days. Data from \citet{maeda08}.
}\label{04gq}}
\end{figure}

\begin{figure}
\centering
\includegraphics[width=\linewidth]{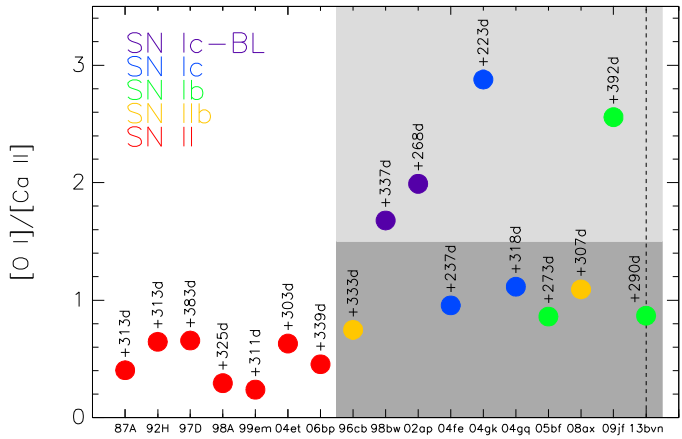}
\caption{[O~I]$\lambda\lambda$6300,6364/[Ca~II]$\lambda\lambda${7291},7324 line ratio of several core-collapse SNe of similar ages during nebular phase.
{Red symbols indicate SN type-II, orange type-IIb, green type-Ib, blue type-Ic, and purple broad-lined type-Ic.} 
{The SN phases with respect to the time of maximum light are shown next to the each data points and dashed vertical line indicates iPTF13bvn. The light-grey shaded region indicates the single star progenitors and the dark-grey shaded region indicates the binary progenitors, for type-Ib/c SNe. The exact border between these two regions is unknown, thus arbitrarily taken at [O~I]/[Ca~II] = 1.5 in this plot for the purpose of indicating the two distinct populations.
References for the spectra of individual SNe: 1987A \citep{pun95}; 1992H \citep{clocchiatti96}; 1997D \citep{benetti01}; 1998A \citep{pastorello05}; 1998bw \citep{patat01}; 1999em \citep{leonard02}; 2004et \citep{sahu06}; 2004gq \citep{maeda08}; 2005bf \citep{maeda07}; 2006bp \citep{quimby07}; 1996cb, 2002ap, 2004fe, 2004gk, 2008ax, 2009jf \citep{modjaz14}.
\label{ratio}}
}
\end{figure}

We also measured the line ratio of [O~I]$\lambda\lambda$6300, 6364/[Ca~II]$\lambda\lambda${{7291}}, 7324 emissions in the nebular spectrum of iPTF13bvn, and compared it to other stripped-envelope SNe as well as several type-II SNe taken from the SUSPECT\footnote{\url{http://www.nhn.ou.edu/~suspect/}} Online Supernova Spectrum Archive. 
{This line ratio is known to be insensitive to the density and temperature, while increases with increasing progenitor mass (\citealt{fransson89}; also see \citealt{elmhamdi04}). The evolution of the ratio of this line with SN age is generally small, as shown in an example of SN 2004gq in Figure~\ref{04gq}. The ratio of [O~I]$\lambda\lambda$6300, 6364/[Ca~II]$\lambda\lambda${{7291}}, 7324 changes from around unity at phase +258 days to 1.3 at +381 days, i.e. only $\sim$30 $\%$ change in more than 120 days. It is also has been shown by \citet[][see their Figure 3]{elmhamdi04} that the ratio is almost constant for SNe aged 280 -- 400 days after the explosion.
}
Figure~\ref{ratio} shows the measured line ratio for iPTF13bvn compared to other SNe. It is clear that the SNe II do not exceed line ratio of 0.7, while on the other hand there is considerable spread for the type-Ib/Ic SNe. 
While the evolution of the line ratio may contribute to the scatter in the diagram, it is considerably insignificant compared to the observed spread\footnote{{We, however, again caution the reader of the caveat that it is possible that this pattern observed in Figure~\ref{ratio} may change with time as different SNe may exhibit different rates of evolution, i.e. some objects may evolve more dramatically compared to the others. Note that this late-time behaviour in SNe is still not very well observed and studied.}}.
This spread may be interpreted as indicative of the presence of two different populations of SNe Ib/c (shaded regions in Figure~\ref{ratio}), i.e. those coming from massive single Wolf-Rayet progenitors, and lower-mass progenitors in binary systems \citep{hk13a}.
In this case, iPTF13bvn appears to be more associated to the lower-mass progenitors of type-II SNe, further strengthening the notion of its progenitor in a binary system. It has been suggested that the majority ($\gtrsim$70 \%) of massive stars undergo binary interaction at some point in their evolutions, thus implying that a significant fraction of core-collapse SN progenitors must experience such phase as well \citep{sana12}.
The confirmation that {the progenitor of iPTF13bvn} indeed belonged to a binary system may eventually be provided by the detection of the remaining companion by future observations when the SN has faded considerably, such as the case of SN IIb 2011dh \citep{folatelli14}. 

\subsubsection{Possible presence of H$\alpha$?} 
\label{sec:ha}
As can bee seen in Figure~\ref{specs}, the global appearance of the late-time spectrum of iPTF13bvn closely resembles that of SN 2008ax. \citet{taubenberger11} reported the presence of a broad feature redward of the [O~I]$\lambda\lambda$6300, 6364 emission line in nebular spectra of SN 2008ax. This feature appeared after $\sim$100 days after the explosion and remained visible up until at least phase 358 days. In other stripped-envelope SNe this feature has also been observed, such as in SN IIb 1993J \citep{patat95} and SN Ib 1996N \citep{sollerman98}. 
{In the nebular spectra of SN Ib 2005bf \citep{maeda07}, a broad feature at $\sim$6500 \AA{ } has been identified as H$\alpha$ (see Fig.~\ref{specs}).}
Also in the case of SN 2010as, a low-velocity SN IIb showing Ib/c characteristics, this feature was also observed \citep{folatelli14as}. During the transitional phase from photospheric to nebular at around +100 days past maximum light, the profile is complex, possibly indicating the presence of other lines, and appears to be still present well into the nebular phase at +309 days.

\begin{figure}
\centering
\includegraphics[width=\linewidth]{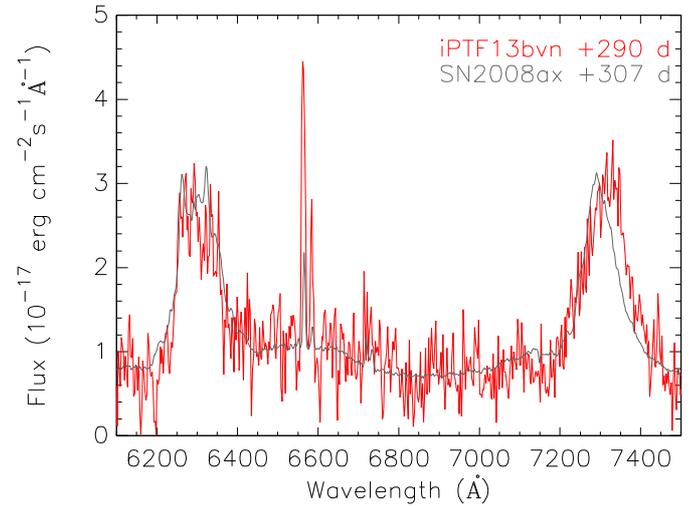}
\caption{Spectra of iPTF13bvn and SN 2008ax around H$\alpha$ region. The flux of the SN 2008ax spectrum has been normalized and shifted.\label{specha}}
\end{figure}

Although it is compelling to attribute this feature to late-time H$\alpha$ emission from the SN, as has been discussed by \citet{taubenberger11} there could be other possible sources of this broad feature such as various other elements including Fe, and scattering. In Figure~\ref{specha} we plot the spectrum of iPTF13bvn alongside SN 2008ax from \citet[][same data shown in Figure~\ref{specs}]{modjaz14} around the H$\alpha$ region.  The graph is suggestive of the presence of a broad feature redward of the [O~I] line in the iPTF13bvn spectrum, very similar to SN 2008ax. However, due to the possible ambiguity in the identification as discussed above and low signal-to-noise ratio in the spectrum we restrain ourselves from firmly associating this feature with H$\alpha$ emission from the SN itself. Furthermore, \citet{jerkstrand15} argued that the thin hydrogen envelope in type-IIb SNe should not affect the spectrum after around 150 days. They attribute the feature redward of the [O~I] line in the nebular spectra of SNe 1993J, 2008ax, and 2001dh to emission from [N~II]$\lambda\lambda$6548, 6583. 

\begin{figure*}
\centering
\begin{subfigure}{.5\textwidth}
  \centering
  \includegraphics[width=\linewidth, height=6.5cm]{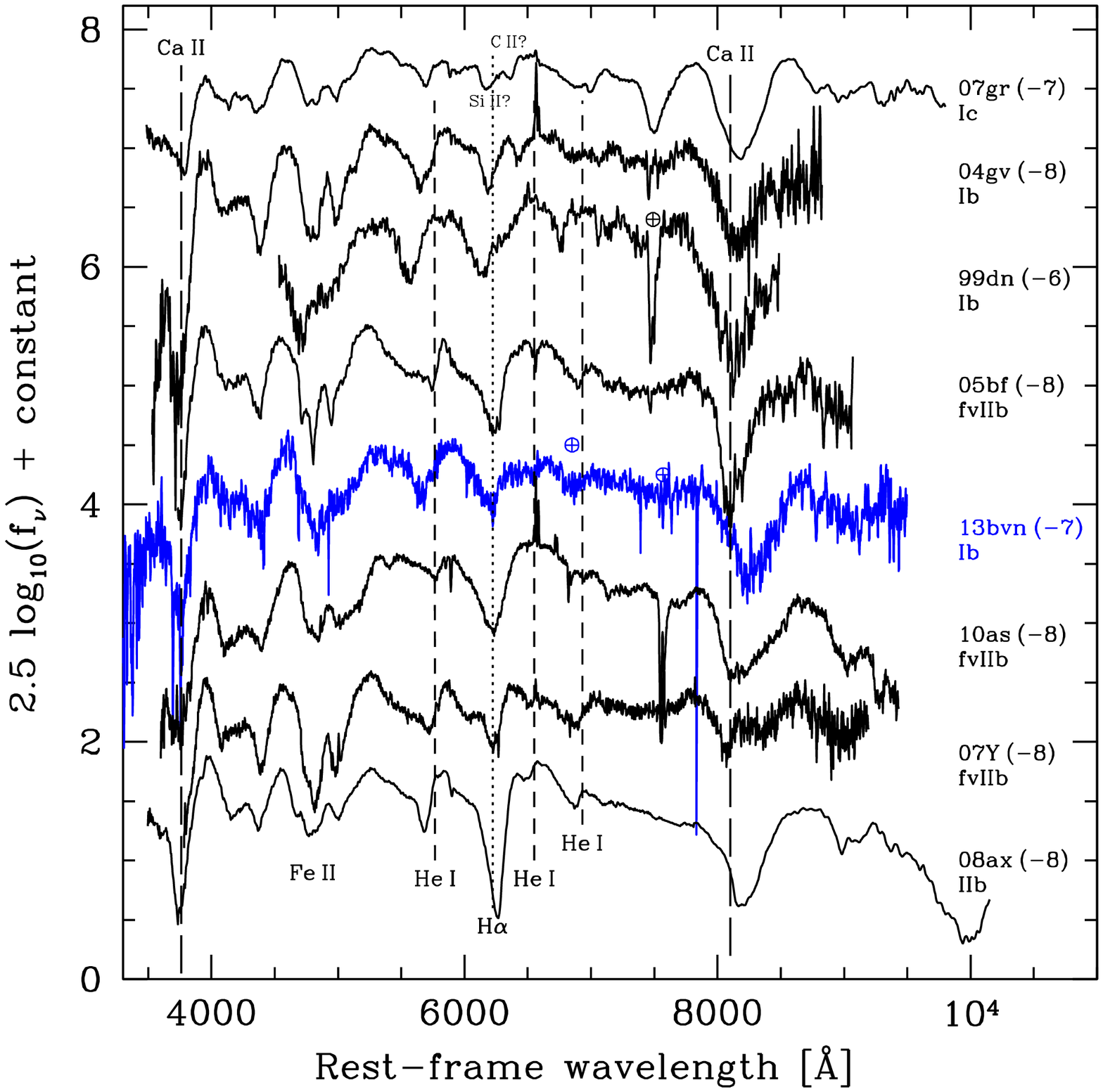}
\end{subfigure}%
\begin{subfigure}{.5\textwidth}
  \centering
  \includegraphics[width=\linewidth, height=6.5cm]{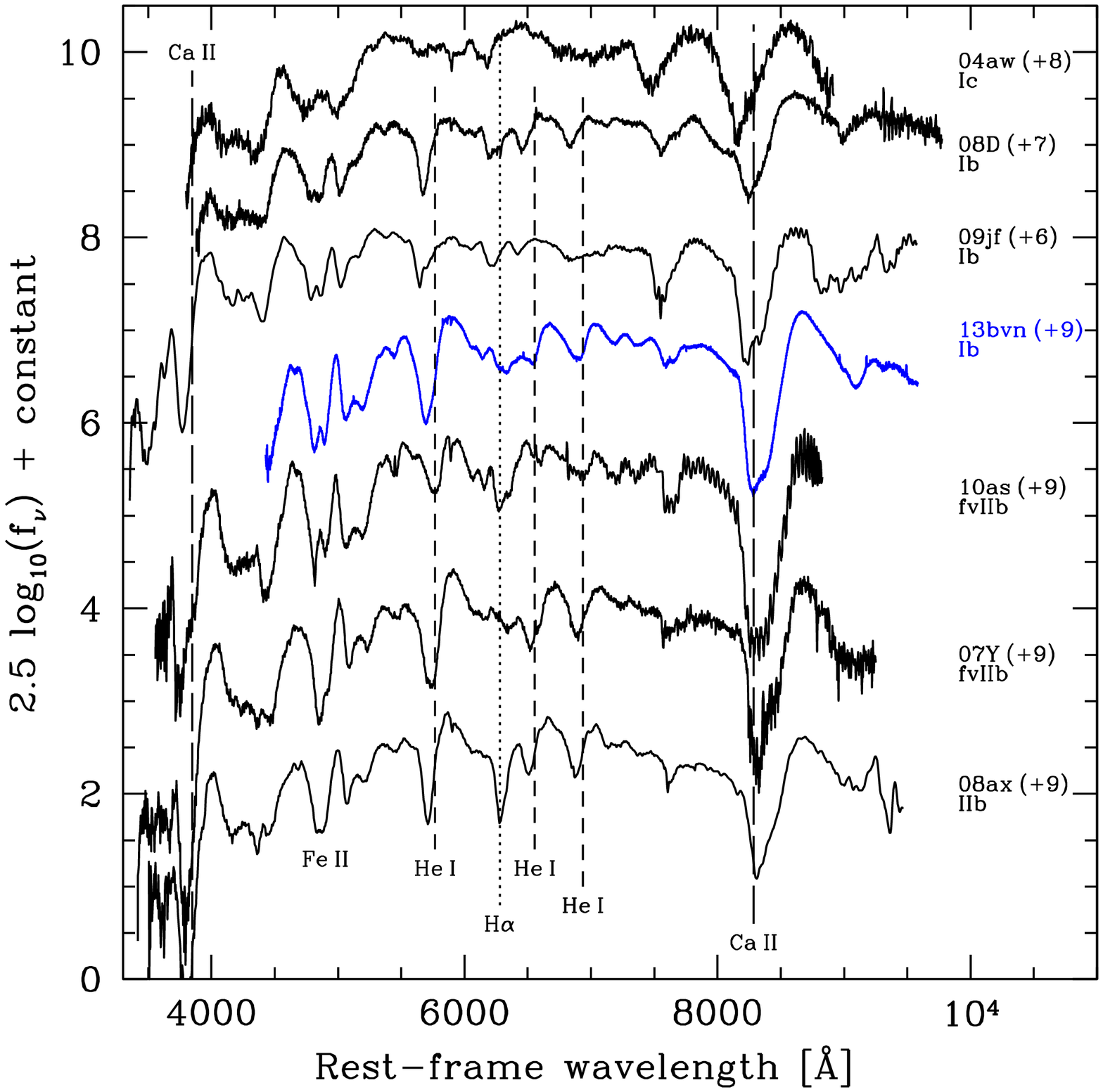}
\end{subfigure}
\caption{Early spectra of iPTF13bvn at around one week before and after maximum light, compared to other stripped-envelope SNe. Vertical dashed lines show the positions of prominent helium lines, and dotted line show the position of high-velocity H$\alpha$. The spectra were taken from the Weizmann Interactive Supernova Data Repository \citep[WISeREP;][]{yaron12}
\label{earlyspec}}
\end{figure*}

We note that the early-phase spectra of iPTF13bvn bear similarities with SNe 2007Y and 2010as, which were shown to belong to the class of "flat-velocity type-IIb" transitional SNe \citep{folatelli14as}. These objects show SN type-Ib/c signatures in their spectra, but also the presence of hydrogen, and characteristic flat velocity evolution between 6000 and 8000 km~s$^{-1}$. The velocity evolution of iPTF13bvn, however, does not show this characteristic flat evolution \citep[see Figure~9 of][]{srivastav14}.
In Figure~\ref{earlyspec} we show the early spectra of 13bvn, around one week before and after maximum light compared to other SNe including 2007Y and 2010as. 
The absorption trough at 6200~\AA~ in iPTF13bvn has been attributed to possibly Si~II or Ne~I by \citet{srivastav14} and \citet{cao13}. In the case of SN 2010as, it was shown that this absorption is best matched with high-velocity H$\alpha$ rather than Si~II \citep{folatelli14as}. 

\subsubsection{H II region metallicity}
\label{sec:metal}
In the spectrum of iPTF13bvn there are narrow emission lines. The emission lines of H$\alpha$, [N~II]$\lambda$6583, and [S~II]$\lambda\lambda${6716, 6731} are presumably produced by the underlying H II region at the SN position. Measurement of the logarithm of the flux ratio of [N II] to H$\alpha$ emission lines, the \textit{N2} index, can be used as an indicator of metallicity. \citet{pp04} gives the oxygen abundance as $12+\log\textrm{(O/H)}=8.90+0.57 \times N2$. Our measurement of the \textit{N2} index shows that the H II region associated with iPTF13bvn has the gas-phase metallicity of 12~+~log(O/H)~=~8.63. This suggests that the metallicity of the progenitor of iPTF13bvn is close to the solar value of 12~+~log(O/H)~=~8.69 \citep{asplund09}, i.e. about 0.87 Z$_\odot$. 

We also measured the oxygen abundance from a nearby H II region 0.7 arcsec away from the SN, corresponding to projected physical distance of 89 pc. We used the data taken as part of the Carnegie Supernova Project \citep[CSP;][]{hamuy06}, using the 6.5 m Magellan Baade telescope at Las Campanas Observatory, Chile, and the instrument IMACS \citep{bigelow98} on longslit spectroscopy mode with 300 lines/mm grating blazed at 4.3$^\circ$. {The} observation was carried out in the night of 13 August 2014 (local time), under 0.5 arcsec seeing. Measurement of the \textit{N2} index yields 12 + log(O/H) = 8.62, which is very close to the derived on-site metallicity of 8.63. 

As already mentioned in subsection \ref{sec:oxmass}, at solar metallicity the minimum initial mass for a single star to become a Wolf-Rayet star via stellar wind is $\sim$25 M$_\odot$ \citep{crowther07}. Considering the metallicity estimate and core oxygen mass, it is unlikely that iPTF13bvn was produced by a $\gtrsim$25 M$_\odot$ progenitor.

\subsection{Late-time light curve}
\begin{figure}
\centering
\includegraphics[width=\linewidth]{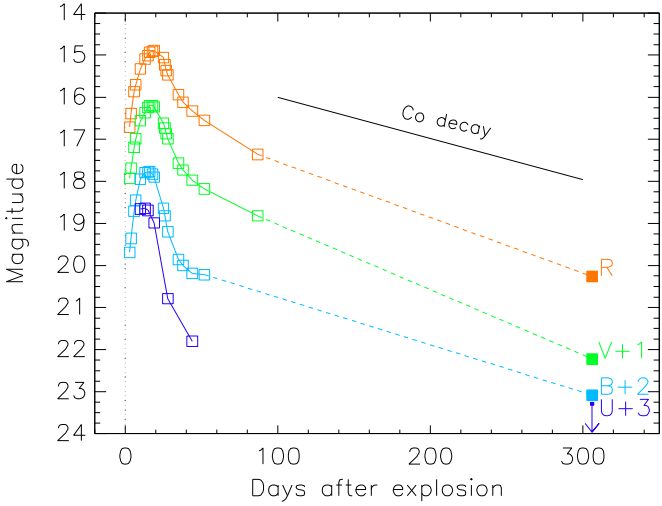}
\caption{Light curve of iPTF13bvn up until day 306
{after the explosion. Explosion time is indicated with dotted vertical line.}
Early photometric data from \citet{srivastav14} {are plotted in open squares}; the latest points in the early light curves are connected to our data points ({filled squares}) at nebular phase by dashed lines.
Typical photometric error bars are smaller than the plotting symbol size.\label{lc}}
\end{figure}

Our photometry unfortunately sampled only one epoch during the late-time evolution of iPTF13bvn. In Figure~\ref{lc} we plot our \textit{UBVRc} data points at 306 days together with the light curve constructed using published photometric data of \citet{srivastav14}. Assuming a constant decay rate since the latest points in the early light curves (+87 days in \textit{R} and \textit{V}, +52 days in \textit{B}), the late-time tails of the light curves appear to decay more rapidly compared to the $^{56}$Co decay rate of 0.98 mag (100 d)$^{-1}$ \citep{woosley89}, i.e. 1.32 mag, 1.55 mag, and 1.13 mag (100 d)$^{-1}$ in \textit{R}, \textit{V}, and \textit{B} bands respectively.

The steeper light curve decay compared to $^{56}$Co signifies that complete $\gamma$-ray trapping did not occur in the case of iPTF13bvn. This is a common phenomenon observed in other stripped-envelope SNe during the nebular phase e.g. SN 2008ax \citep{taubenberger11}, SN 1996N \citep{sollerman98}, SN 1990I \citep{elmhamdi04}, and is associated with relatively low ejecta mass.
The decline of the light curve of SN 1990I in nebular phase is the steepest in the \textit{V}-band, followed by \textit{R} and then \textit{B}, which is likely also the case in iPTF13bvn although the lack of data points during the late-time evolution of the light curve prevents us from drawing such conclusion. 

\section{Summary}

We present late-time photometric and spectroscopic observations of the type-Ib SN iPTF13bvn. A progenitor candidate of this SN has been detected in pre-explosion HST images, which is potentially the first for type Ib/c SNe. Our spectroscopy shows that the nebular spectrum of the SN is dominated by emission lines of Mg~I], [O~I], and [Ca~II]. 
{Measurement of the flux of [O~I]$\lambda\lambda$6300, 6464 line shows that the oxygen mass of iPTF13bvn does not significantly exceed $\sim$0.7 M$_\odot$, assuming that the emission is produced by 50$\%$ of the total oxygen mass, suggesting that the progenitor star could not have been more massive than $\sim$15 -- 17~M$_\odot$. The uncertainty originated from distance estimate and oxygen line flux measurements are negligible compared to the error from temperature and unseen oxygen.}
Comparing the nebular [O~I]/[Ca~II] line ratio with other SNe, iPTF13bvn appears to be associated with low-mass progenitor stars.

This piece of evidence provides another support to the scenario in which iPTF13bvn was produced by a hydrogen-poor progenitor star in a binary system that has undergone envelope stripping by means of close binary interaction. The initial mass of the progenitor star was not sufficiently high for the star to have become a Wolf-Rayet star, were it a single star. 
{This result shows very good agreement with the results from other methods in constraining the progenitor star, such as from hydrodynamical modeling \citep{bersten14}, analytical modeling \citep{srivastav14}, and binary evolution modelling \citep{bersten14,eldridge15}. In the context of hydrogen-poor SNe, iPTF13bvn further reinforces the importance of massive close binaries as a prominent, if not the dominant, progenitors of these SNe.}

\begin{acknowledgements}
{We thank the anonymous referee for his/her useful comments that substantially improved this paper.}
Support for HK, SGG, MH, TdJ, and CPG is provided by the Ministry of Economy, Development, and Tourism's Millennium Science Initiative through grant IC120009, awarded to The Millennium Institute of Astrophysics, MAS. HK and SGG acknowledge support by CONICYT through FONDECYT grants 3140563 and 3130680, respectively. HK also acknowledges helpful support from staffs of Cerro Pach\'on and NOAO South during the observing run.
KM acknowledges financial support by Grant-in-Aid for Scientific Research (No. 23740141 and 26800100) from the Japanese Ministry of Education, Culture, Sports, Science and Technology (MEXT). The work by KM is partly supported by WPI Initiative, MEXT, Japan.
This research has made use of the NASA/IPAC Extragalactic Database (NED) which is operated by the Jet Propulsion Laboratory, California Institute of Technology, under contract with the National Aeronautics and Space Administration. 
\end{acknowledgements}



%
%
%
%
%
%
%
%
%

\end{document}